# Demonstrating an Improved Length-weight Model in Largemouth Bass, Chain Pickerel, Yellow Perch, Black Crappie, and Brown Bullhead in Stilwell Reservoir, West Point, New York

Mercedes Dexter, Kyle Van Alstine, and Michael Courtney
U.S. Air Force Academy, 2354 Fairchild Drive, USAF Academy, CO 80840
Michael.Courtney@usafa.edu

Ya'el Courtney
BTG Research, PO Box 62541, Colorado Springs, CO 80962

**Background:** The traditional power law model, $W(L) = aL^b$, is widely applied to describe weight (W) vs. length (L) in fish. The exponent, b, is independent of the system of units and has an easily interpreted physical meaning as related to isometric growth (b = 3). In contrast, the coefficient, a, depends strongly on both the exponent and the system of units, and its physical meaning is difficult to interpret. It has been suggested that unit conversions and lack of a physical meaning may have contributed to errors in length-weight parameters at FishBase.org and other widely cited sources.

**Materials and Methods:** The model, $W(L) = (L/L_1)^b$, is proposed as an improvement. The Levenberg-Marquardt non-linear least squares technique is used to determine the best-fit parameters $L_1$ and b. This model has the advantages that $L_1$ has the same units (length) independent of the value of the exponent and has an easily interpreted physical meaning as the typical length of a fish with one unit of weight. This proposed model is compared with the traditional model on length-weight data sets for black crappie, largemouth bass, chain pickerel, yellow perch, and brown bullhead obtained from Stilwell Reservoir, West Point, New York. The resulting best-fit parameters, parameter standard errors, and covariances are compared between the two models. The average relative weight for these species is determined, along with typical meat yields for four species.

**Results:** For the five species, using the logarithmic approach and a linear least-squares, standard errors in the coefficient, a, range from 60.2% to 136.5% for the traditional model. Using a non-linear least squares technique to determine best fit parameters, the standard errors for the coefficient, a, range from 68.5% to 164.0% in the traditional model. In the improved model, standard errors in the parameter $L_1$ range from 0.94% to 15.0%. The covariance between a and b in the traditional model has a magnitude between 0.999 and 1.000 in both linear and non-linear parameter estimation methods. In the improved model, the covariances between $L_1$ and b are smaller. The average relative weights for each species were black crappie (92.0%), largemouth bass (91.8%), yellow perch (69.3%), chain pickerel (88.1%), and brown bullhead (100.4%). The average meat yield in fillets as a percentage of total weight were largemouth bass (34.5%), yellow perch (34.9%), chain pickerel (41.9%), and brown bullhead (22.5%).

**Conclusion:** In this study, the improved model, $W(L) = (L/L_1)^b$, is preferable for weight vs. length in fish, because the estimated parameter uncertainties and covariances are smaller in magnitude. Furthermore, the parameters both have consistent units and an easily interpreted physical meaning.

## I. Introduction

The traditional power law model, $W(L) = aL^b$, finds widespread application for length-weight relationships in fish. Taking the logarithm of both sides in this equation allows estimation of best-fit parameters by means of linear least-squares regression. (Anderson and Neumann 1996) Alternatively, the parameters can be estimated by the Levenberg-Marquardt non-linear least-squares method. The exponent, b, in this model is independent of the system of units and has an easily interpreted physical meaning as related to isometric growth for b = 3. (Pauly 1984) In contrast, the coefficient, a, depends strongly on both the exponent and units, and its physical meaning is difficult to interpret. It has been suggested that unit conversions and lack of a clear physical meaning may have contributed to widespread errors in length-weight parameters at FishBase.org.(Cole-Fletcher 2011)

The alternate model, $W(L) = (L/L_1)^b$, is proposed as an improvement. The Levenberg-Marquardt non-linear least-squares technique is used to determine the best-fit parameters $L_1$ and b. This model has

# Demonstrating an Improved Length-weight Model in Largemouth Bass, Chain Pickerel, Yellow Perch, Black Crappie, and Brown Bullhead in Stilwell Reservoir, West Point, New York

the advantages that $L_1$ has the same units (length) independent of the value of the exponent and has an easily interpreted physical meaning as the typical length of a fish with one unit of weight. This proposed model is compared with the commonly used model on length-weight data sets for black crappie (*Pomoxis nigromaculatus*), largemouth bass (*Micropterus salmoides*), chain pickerel (*Esox niger*), yellow perch (*Perca flavescens*), and brown bullhead (*Ameiurus nebulosus*) obtained from Stilwell Reservoir, West Point, New York. The resulting best-fit parameters, parameter standard errors, and covariances are compared between the two models. Furthermore, the average relative weight for these species is determined. (Anderson and Neumann 1996) Typical meat yields are also determined for four species.

Stilwell Reservoir is a clear mountain reservoir located in the Hudson Valley in Orange County, New York on the West Point Military Reservation. The reservoir has a surface area of 0.52 km$^2$, and a 15 m maximum depth. The average depth is 6 m. Fishing is restricted to United States military personnel and Department of Defense civilian employees. Mr. James Beemer, the United States Military Academy's fish and wildlife biologist, maintains Stilwell Reservoir and keeps it well stocked with various kinds of fish.

## II. Method

Fish samples were obtained by sport angling. Fish were relatively easy to catch at rates of 2-5 fish per hour by trolling crank baits at 1-2 km per hour behind a boat and by fishing using small bait fish ("shiners" and small "sunfish" caught at the site) 1-5 m below a bobber. The fish were then weighed and measured from the foremost lip to the fork in the tail.

Most weight vs. length models in fish are in the form of a traditional power law function, $W(L) = aL^b$, where W is the weight of the fish and L is the fork length (FL) or total length (TL). Here, an improved model, $W(L) = (L/L_1)^b$ is used and the best-fit parameters are determined with the Levenberg-Marquardt non-linear least-squares algorithm. This approach has the advantages of equal weighting (or more easily controlled weighting) for all the data points in the sample, smaller covariance between the two model parameters, and an easily interpreted physical meaning. The parameter $L_1$ has the same units of length as used in the model, and its physical meaning is the typical length of the fish that has one unit of weight. If weight is measured in kg, then $L_1$ is the typical length of a fish that weights 1 kg. If desired, the equivalent parameter a can then obtained by simple algebra.

Fish were also compared with the standard weight curve generated from length-weight parameters in Anderson and Neumann (1996) or Bister et al. (2000). This comparison requires converting total length (TL) in the standard weight curve to fork length (FL) by dividing by an appropriate factor.

Demonstrating an Improved Length-weight Model in Largemouth Bass, Chain Pickerel, Yellow Perch, Black Crappie, and Brown Bullhead in Stilwell Reservoir, West Point, New York### III. Results
1. **Black Crappie (*Pomoxis nigromaculatus*)**

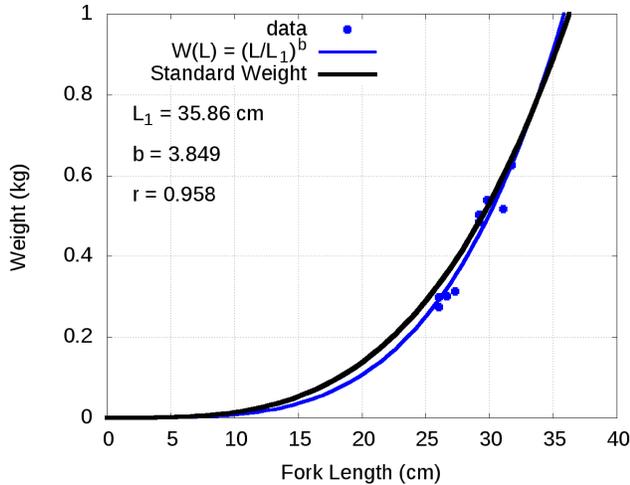

*Figure 1: Weight vs. fork length in black crappie along with best-fit model and standard weight curve. (Anderson and Neumann 1996)*

Judging by the rings on their scales, the black crappie in this sample ranged from approximately three to five years old. The weight vs. length for black crappie is shown in Figure 1. The standard weight curve has been adapted to fork length using FL=TL/1.040. (Froese and Pauly 2010) Using the published standard weight curve, (Anderson and Neumann 1996) the average relative weight was 92.0% +/- 3.2%. Since the fish were preserved whole for another study, fillet yield data was not obtained.

2. **Largemouth Bass (*Micropterus salmoides*)**

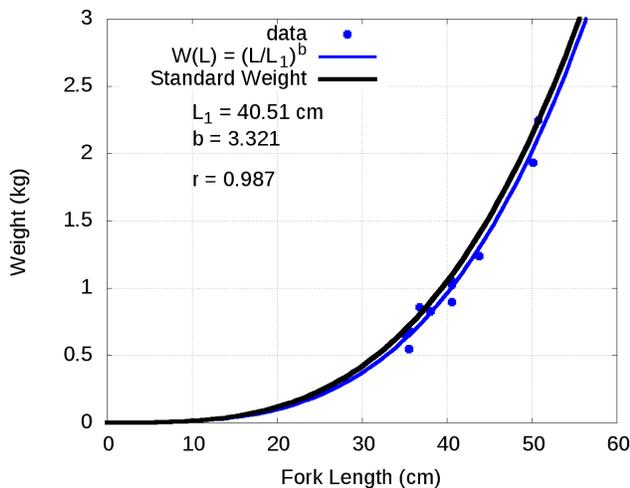

*Figure 2: Weight vs. fork length in largemouth bass along with best-fit model and standard weight curve. (Anderson and Neumann 1996)*

# Demonstrating an Improved Length-weight Model in Largemouth Bass, Chain Pickerel, Yellow Perch, Black Crappie, and Brown Bullhead in Stilwell Reservoir, West Point, New York

The largemouth bass is a popular sport fish in much of the United States. In Stilwell Reservoir, largemouth bass are plentiful and easy to catch. The fish measured here were caught by fishing from the bank with minnows or trolling from a boat with popular crank baits. The largemouth bass population in Stilwell Reservoir not only feed around the bank or around large objects (fallen trees), they also chase prey fish into the deeper parts and more open waters of the reservoir. The age of largemouth bass in this sample ranged from approximately three to eight years old, judging by the rings on their scales.

The weight vs. length for largemouth bass is shown in Figure 2. The standard weight curve has been adapted to fork length with FL=TL/1.024. (Froese and Pauly 2010) Using the standard weight to compute the relative weight of wach sample, (Anderson and Neumann 1996) and the average relative weight was 91.8% +/- 3.3%. A linear regression (zero offset) of fillet yield vs. total weight has a slope of 0.345(8) with r=0.990, suggesting a typical fillet yield of 34.5% independent of total weight.

3. **Yellow Perch (*Perca flavescens*)**

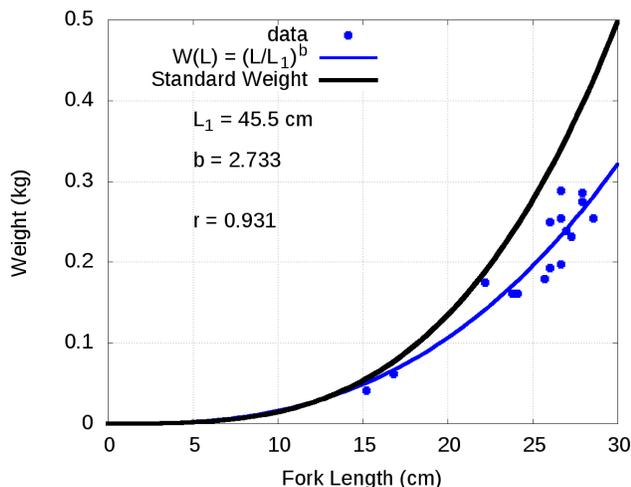

*Figure 3: Weight vs. fork length in yellow perch along with best-fit model and standard weight curve. (Anderson and Neumann 1996)*

The ages of yellow perch caught ranged from approximately two to nine years old, judging by the rings on their scales. The weight vs. length for yellow perch is shown in Figure 3. The standard weight curve has been adapted to fork length with FL=TL/1.060. (Froese and Pauly 2010) The average relative weight was 69.3% +/- 4.0% of the relative weight. (Anderson and Neumann 1996) A linear regression (zero offset) of fillet yield vs. total weight has a slope of 0.349(6) with r=0.958, suggesting a typical fillet yield of 34.9% independent of total weight.

4. **Chain Pickerel (*Esox niger*)**

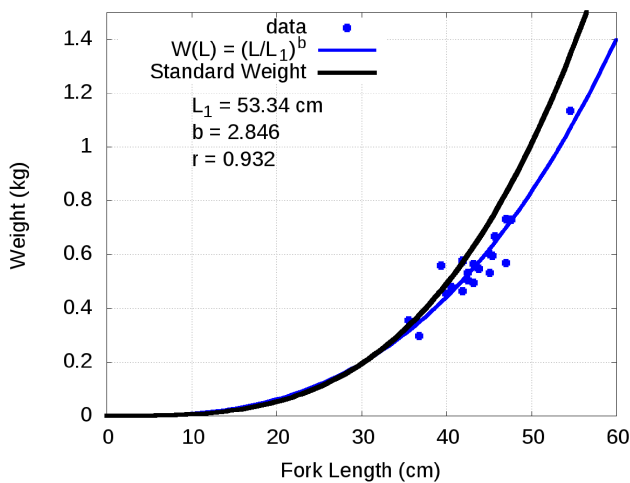

*Figure 4: Weight vs. fork length in chain pickerel along with best-fit model and standard weight curve. (Anderson and Neumann 1996)*

Unlike largemouth bass, the chain pickerel in this sample were rarely caught by trolling. Most were caught fishing from the bank with minnows. Chain pickerel mostly feed within casting distance from the shore, usually near rocky outcroppings. Judging by the rings on their scales, the ages of chain pickerel in this sample are approximately three to six years old. The weight vs. length for chain pickerel is shown in Figure 4. The standard weight curve has been adapted to fork length with FL=TL/1.055. (Froese and Pauly 2010) The average relative weight was 88.1% +/- 2.6% of the standard weight. (Anderson and Neumann 1996) A linear regression (zero offset) of fillet yield vs. total weight has a slope of 0.419(7) with r=0.964, suggesting a typical fillet yield of 41.9% independent of total weight.

5. **Brown Bullhead (*Ameiurus nebulosus*)**

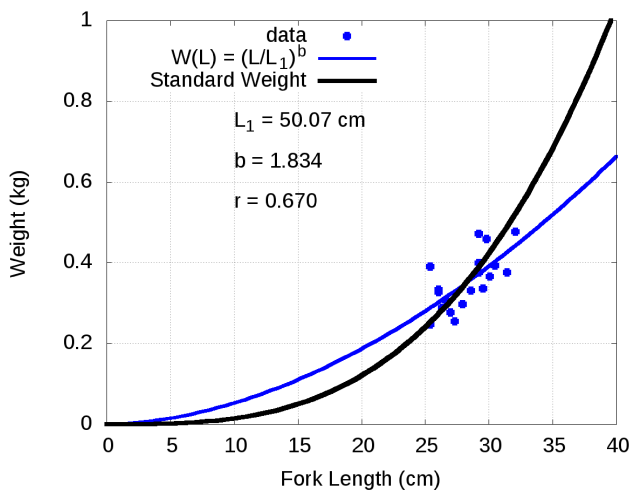

*Figure 5: Weight vs. fork length in brown bullhead along with best-fit model and standard weight curve. (Anderson and Neumann 1996)*

# Demonstrating an Improved Length-weight Model in Largemouth Bass, Chain Pickerel, Yellow Perch, Black Crappie, and Brown Bullhead in Stilwell Reservoir, West Point, New York

Brown bullhead were easily caught near dusk or after dark angling from the bank with small bait fish or the occasional nightcrawler. Age was not determined. The weight vs. length for brown bullhead is shown in Figure 5. The brown bullhead data are not as tightly grouped as other fish. Consequently, the correlation coefficient is r = 0.670, which means the model does not fit the data as well. The best fit parameters have large standard errors; the standard error in the typical length ($L_1$) is 7.5 cm, and the uncertainty in the exponent (b) is 0.49. The cause of the dispersion in the brown bullhead weights is unclear.

The standard weight curve (Bister et al. 2001) has been adapted to fork length with FL=TL/1.008. (Froese and Pauly 2010) The average relative weight was 100.4% +/- 4.4% of the standard weight. (Bister et al. 2001) A linear regression (zero offset) of fillet yield vs. total weight has a slope of 0.225(6) with r = 0.817. The typical fillet yield is 22.5%, and plotting the residual errors suggests despite less uniformity in yield than other species, yield percentage is independent of total weight.

6. **Comparisons of Models and Fitting Method**

Table 1 shows that the parameter $L_1$ in the improved model has units of cm and yields values that are in line with reasonable expectations regarding the length of a 1 kg fish of each type. Standard errors for $L_1$ in the improved model vary from 0.9% (largemouth bass) to 15.0% (brown bullhead). In contrast, the parameter a in the traditional model varies considerably, depending on whether LLS or NLLS estimation methods are used and has much larger standard errors, which vary from 68.5% to 164.0% when a is determined by NLLS and from 60.2% to 136.5% when determined by the LLS after taking the logarithm of both length and weight to obtain a linear equation.

|  | **Best-fit Parameters** |  |  | **Standard** | **Errors** |  |
|---|---|---|---|---|---|---|
|  | NLLS | NLLS | LLS | NLLS | NLLS | LLS |
| **Species** | $L_1$ (cm) | a | a | $L_1$ | a | a |
| **Black Crappie** | 35.862 | 1.040E-6 | 4.732E-7 | 2.4% | 155.0% | 136.5% |
| **Largemouth Bass** | 40.513 | 4.588E-6 | 6.362E-6 | 0.9% | 68.5% | 88.7% |
| **Yellow Perch** | 45.472 | 2.944E-5 | 1.421E-5 | 8.8% | 143.7% | 60.2% |
| **Chain Pickerel** | 53.336 | 1.214E-5 | 2.999E-5 | 1.5% | 87.7% | 103.2% |
| **Brown Bullhead** | 50.074 | 7.650E-4 | 8.625E-4 | 15.0% | 164.0% | 113.4% |

*Table 1: Best-fit parameters a and $L_1$ determined via linear least squares (LLS) and nonlinear least squares (NLLS) fitting of models along with standard errors reported as percentages. The standard errors for the proposed improved model, $W(L) = (L/L_1)^b$ are much smaller.*

Best-fit values of the parameter b depend on whether the estimation is LLS or NLLS, as shown in Table 2. The covariance between parameters a and b is very close to -1 in the traditional length-weight model, showing a strong relationship between the two: decreasing one parameter significantly increases the other for the model to remain close to the data. Covariances between parameters b and $L_1$ are all greater in magnitude than 0.5, also showing strong dependence, though not as close to 1 as in the traditional model.

Demonstrating an Improved Length-weight Model in Largemouth Bass, Chain Pickerel, Yellow Perch, Black Crappie, and Brown Bullhead in Stilwell Reservoir, West Point, New York

|  | Parameters | | Standard Errors | | Covariance | |
|---|---|---|---|---|---|---|
|  | NLLS | LLS | NLLS | LLS | NLLS | NLLS/LLS |
| **Species** | b | b | b | b | Improved | Traditional |
| **Black Crappie** | 3.849 | 4.081 | 0.4614 | 0.4072 | 0.947 | -0.999 |
| **Largemouth Bass** | 3.321 | 3.233 | 0.1804 | 0.2398 | 0.663 | -0.999 |
| **Yellow Perch** | 2.733 | 2.955 | 0.4407 | 0.1876 | 0.991 | -0.999 |
| **Chain Pickerel** | 2.846 | 2.606 | 0.2311 | 0.2741 | 0.851 | -0.999 |
| **Brown Bullhead** | 1.834 | 1.795 | 0.4903 | 0.7143 | 0.992 | -0.999 |

*Table 2: Best-fit parameter b determined via linear least squares (LLS) and nonlinear least squares (NLLS) fitting of models along with standard errors. Covariances are also compared. The traditional model, $W(L) = aL^b$, and the improved model, $W(L) = (L/L_1)^b$, both yield the same values for b when estimated by the NLLS method. However, the traditional model yields different values for the exponent b when estimated by the LLS method after taking the logarithm of length and weight.*

## IV. Discussion

In the test cases considered here, the improved length-weight model produces realistic values for the typical length parameter, smaller magnitude covariances between parameters, and much smaller standard errors in the non-exponent parameter. The traditional power law length-weight model seems like an artifact from a time before widespread implementation of NLLS algorithms on digital computers. Many papers using the traditional power law model fail to describe whether parameters are determined by LLS or NLLS methods, even though these can yield different parameter estimates. Many papers also fail to report standard errors in the parameters a and b. The improved model, $W(L) = (L/L_1)^b$, is easily generalizable to grams and millimeters and also to cases where having the length scale parameter correspond to a different typical weight is more appropriate. For example, $W(L) = 10(L/L_{10})^b$, $W(L) = 100(L/L_{100})^b$, and $W(L) = 1000(L/L_{1000})^b$, could be used for species where it makes the most sense to have a typical length parameter correspond to a fish weighing 10 g, 100 g, and 1000 g, respectively.

One could argue that fisheries science is doing fine describing length-weight relationships with the traditional model. However, recently discovered errors in length-weight parameters at FishBase.org (Cole-Fletcher et al. 2011), an on-line database of fish related data with over 1500 citations, suggest some benefit to a typical length parameter where absurd values can be spotted by simple inspection because of its clear physical meaning.

## V. Acknowledgements

The authors acknowledge helpful discussions with Amy Courtney, PhD (BTG Research) and Beth Schaubroeck, PhD (USAFA Department of Mathematical Sciences). We thank the United States Military Academy at West Point for graciously allowing us to fish in Stilwell Reservoir to collect this data and the Quantitative Reasoning Center at the United States Air Force Academy and BTG Research for supporting this work. Peer-reviewers also provided helpful suggestions.

# Demonstrating an Improved Length-weight Model in Largemouth Bass, Chain Pickerel, Yellow Perch, Black Crappie, and Brown Bullhead in Stilwell Reservoir, West Point, New York